\documentclass[twocolumn,showpacs,preprintnumbers,amsmath,amssymb,prl]{revtex4}
\usepackage{graphicx}% Include figure files
\usepackage{subfigure}
\usepackage{dcolumn}% Align table columns on decimal point
\usepackage{bm}% bold math
\usepackage{epstopdf}
\begin{document}

\title{Experimental determination of the Weiss temperature of Mn$_{12}$-ac and Mn$_{12}$-ac-MeOH}
\author{Shiqi Li}
\affiliation{Department of Physics, City College of New York, CUNY, New York, New York
10031, USA}
\author{Lin Bo}
\affiliation{Department of Physics, City College of New York, CUNY, New York, New York
10031, USA}
\author{Bo Wen}
\affiliation{Department of Physics, City College of New York, CUNY, New York, New York
10031, USA}
\author{P. Subedi}
\affiliation{Department of Physics, New York University, New York, New York 10003, USA}
\author{Y. Yeshurun}
\affiliation{Department of Physics, City College of New York, CUNY, New York, New York
10031, USA}
\affiliation{Department of Physics, New York University, New York, New York 10003, USA}
\affiliation{Department of Physics, Institute of Nanotechnology, Bar-Ilan University,
Ramat-Gan 52900, Israel}
\author{A. D. Kent}
\affiliation{Department of Physics, New York University, New York, New York 10003, USA}
\author{M. P. Sarachik}
\affiliation{Department of Physics, City College of New York, CUNY, New York, New York
10031, USA}
\author{A. J. Millis}
\affiliation{Department of Physics, Columbia University, New York, New York 10027, USA}
\author{C. Lampropoulos}
\author{S. Mukherjee}
\author{G. Christou}
\affiliation{Department of Chemistry, University of Florida, Gainesville, Florida 32611,
USA}

\date{April 30, 2010}

\begin{abstract}
We report measurements of the susceptibility in the temperature range from $3.5$ K to $6.0$ K of a series of Mn$_{12}$-ac and Mn$_{12}$-ac-MeOH samples in the shape of rectangular prisms of length $l_c$ and square cross-section of side $l_a$.  The susceptibility obeys a Curie-Weiss Law, $\chi=C/(T-\theta)$, where $\theta$ varies systematically with sample aspect ratio. Using published demagnetization factors, we obtain $\theta$ for an infinitely long sample corresponding to intrinsic ordering temperatures $T_c \approx 0.85$ K and $\approx 0.74$ K for  Mn$_{12}$-ac and
Mn$_{12}$-ac-MeOH, respectively.  The difference in $T_c$ for two materials that have nearly identical unit cell volumes and lattice constant ratios suggests that, in addition to dipolar interactions, there is a non-dipolar (exchange) contribution to the Weiss temperature that differs in the two materials because of the difference in ligand molecules.
\end{abstract}

\maketitle

\section{Introduction}

The magnetic susceptibility, $\chi$, is given within the mean-field approximation by the Curie-Weiss law, $\chi = C/(T - T_c)$, where $C$ is a material-specific Curie constant, $T$ is the absolute temperature and the Weiss temperature $T_c$ is the (approximate) temperature below which the system is ferro- or antiferromagnetic, depending on the sign of $T_c$.  In order to determine the susceptibility and $T_c$ experimentally, however, the measured values need to be corrected to account for a demagnetizing field, $H_d$, which depends on sample geometry, as well as the value of the susceptibility itself.

In this paper we report measurements of the magnetization and susceptibility of a series of samples of two different variants of the molecular magnet Mn$_{12}$-ac: the usual, much-studied form referred to as Mn$_{12}$-ac and a new form abbreviated as Mn$_{12}$-ac-MeOH.  The measurements were taken for a series of samples in the shape of a rectangular prisms with approximately square cross-section  $l_a^2$ and length $l_c$.  The susceptibility is found to obey the expected Curie-Weiss form, $\chi = C/(T - \theta)$ \cite{Luis,Wen}, but
with a Curie-Weiss $\theta$ that varies systematically with the sample
aspect ratio $l_c/l_a$.  Using demagnetization factors calculated by
Chen, Pardo, and Sanchez \cite{Chen2004,Chen2005}, we have deduced the
value of $\theta$ for an infinitely long sample, corresponding to
the intrinsic ordering temperature $T_c$ for the two materials.

\section{Experimental Procedure}

Parallel studies were carried out on single crystals of the usual form of Mn$_{12}$-ac, [Mn$_{12}$O$_{12}$(O$_2$CMe)$_{16}$(H$_2$O)$_4$]$\cdot$2MeCO$_2$H$\cdot$4H$_2$O, and a new recently synthesized form Mn$_{12}$-ac-MeOH, [Mn$_{12}$O$_{12}$(O$_2$CMe)$_{16}$(MeOH)$_4$]$\cdot$MeOH.   The normal form (space group $I\bar{4}$; unit cell parameters $a = b = 17.1668(3)$ \AA, $c = 12.2545(3)$ \AA, $Z = 2, V = 3611.39$ \AA$^3$ at $83$ K) \cite{Cornia} and the new form (space group $I\bar{4}$; unit cell parameters $a = b = 17.3500(18)$ \AA, $c = 11.9971(17)$, $Z = 2, V = 3611.4$ \AA$^3$ at $-100\,^{\circ}\mathrm{C}$) \cite{Stamatatos} of Mn$_{12}$-ac are quite similar, but they differ in the solvent molecules of crystallization that lie in-between the Mn$_{12}$ molecules. In normal Mn$_{12}$-ac, each Mn$_{12}$ molecule forms O-H..O hydrogen-bonds with $n$ $(n = 0-4)$ of the surrounding MeCO$_2$H molecules while in Mn$_{12}$-ac-MeOH, the lattice MeOH molecules form no hydrogen bonds to the Mn$_{12}$ molecules.

Sample preparations for Mn$_{12}$-ac and Mn$_{12}$-ac-MeOH are described in Refs. \cite{Lis} and \cite{Stamatatos}, respectively.  The samples were in the form of rectangular prisms of dimensions $l_a \times l_b\times l_c$ ($l_a \approx l_b$) with
$l_a$ varying from $\sim 0.1$ mm to $\sim0.4$ mm.  The dimensions of the samples were measured under a microscope by a small scaler.  Data were taken for aspect ratios ($l_c/l_a$) varying from $0.75$ to $9.57$ for Mn$_{12}$-ac, and from $1.45$ to $4.9$ for Mn$_{12}$-ac-MeOH.  The range of aspect ratios was determined by sample availability.  

The magnetization was measured in a commercial Quantum Design MPMS (Magnetic Property Measurement System) SQUID-based magnetometer.  The crystals were mounted using a minimum
amount of Dow Corning high vacuum grease.  The Mn$_{12}$-ac-MeOH crystals degrade rapidly when removed from their mother liquor and are very difficult to handle.  These samples were transferred as quickly as possible into paraffin oil using a paraffin-coated stick.  Care was taken to align the
$c$-axis of the crystals parallel to the magnetic field.

The measured magnetization should be normalized by the volume (or mass) of each sample.  In our case, the samples are so small that neither the volume, $V$, nor the mass, $m (\approx 10^{-5}$ g), can be measured accurately.  The SQUID-based magnetometer, however, provides a precise measure of the saturation magnetization, which is proportional to the volume.  We therefore normalized the data for each sample by  $M_{sat}$ and, noting that there are two Mn$_{12}$ molecules, each with spin $S=10$ in a (body-centered cubic) unit cell of known volume, we applied a (calculated) conversion factor $M_{sat} =  2gS\mu_B/V_{cell} =102. 7$ emu/cm$^3$ to obtain the magnetization and the susceptibility in cgs units.

\begin{figure}[h]
\includegraphics[width=0.50\textwidth]{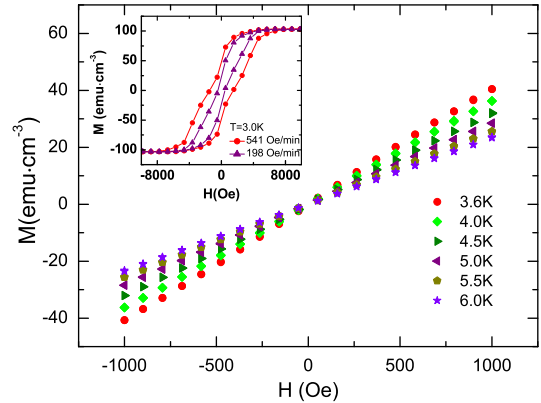}
\caption{\label{hys}(Color online) Magnetization of Mn$_{12}$-ac as a function of
longitudinal magnetic field for different temperatures above the blocking temperature $T_B$.  Sufficiently slow sweep rates are chosen to ensure that the system is in equilibrium.  Inset: open hysteresis loops displaying the familiar steps associated with spin-tunneling in molecular magnets are obtained at
lower temperature and higher sweep rates.}
\end{figure}

Mn$_{12}$-ac molecules behave as nanomagnets with spin $S = 10$ in crystals with strong uniaxial
magnetic anisotropy along the $c$-axis of the crystals.  Modeled as a double-well potential, slow
relaxation below a sweep-rate-dependent blocking temperature $T_B$
gives rise to hysteresis in $M$ versus $H$, as shown in the inset to
Fig. \ref{hys}; the step-wise change of magnetization is typical for
molecular magnets, where steps occur due to spin tunneling at values of longitudinal
magnetic field where energy levels corresponding to different spin
projections cross on opposite sides of the anisotropy barrier.
Equilibrium measurements from which the susceptibility
can be obtained therefore require sufficiently
high temperatures (above blocking) and/or slow sweep rates.
Reversible behavior, signaling equilibrium, was obtained in our
experiments above a blocking temperature on the order of $3$ K to $4$ K
for the sweep rates used.  As a consequence, our susceptibility
measurements were limited to temperatures above $3$ K.

\begin{figure}[h]
\includegraphics[width=0.50\textwidth]{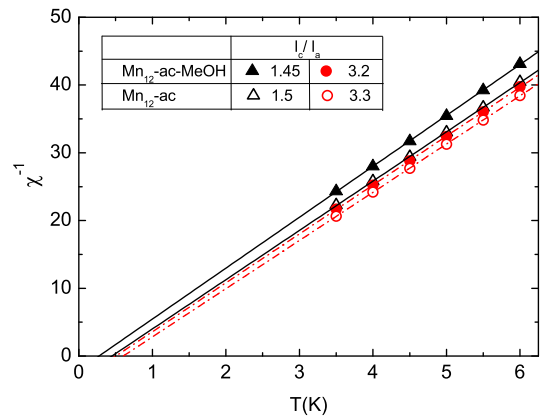}
\caption{\label{slope}(Color online) The inverse susceptibility of Mn$_{12}$-ac as a function of temperature for two samples each of Mn$_{12}$-ac and Mn$_{12}$-ac-MeOH with different (matched)
aspect ratios, as indicated.  The dot-dashed and solid lines are for aspect ratios $\approx 1.5$ and $\approx 3.3$, respectively.  Extension of the straight lines yield temperature intercepts, $\theta$, that vary with aspect ratio and are different for the two materials.}
\end{figure}

\section{Experimental Results}

As shown in Fig. \ref{hys}, data for the magnetization $M$ versus $H$ were obtained in the linear regime.  The susceptibility given by the slope of these straight lines, $\chi=dM/dH|_{H=0}$, increases with decreasing temperature, as expected.

Figure \ref{slope} shows the inverse of the susceptibility as a function of temperature for two crystals of Mn$_{12}$-ac with aspect ratios $l_c/l_a=1.5$ and $3.3$, and two samples of Mn$_{12}$-ac-MeOH with aspect ratios closely matched to those of Mn$_{12}$-ac.  The straight lines demonstrate that the susceptibility obeys a Curie-Weiss law, $\chi=C/(T-\theta)$.  The lines are approximately parallel, indicating that the Curie constant, $C=0.138$ for the two systems are approximately the same, as expected.  For each material, (Mn$_{12}$-ac or Mn$_{12}$-ac-MeOH), the intercept $\theta$ is larger for the larger aspect ratio.  A cross-comparison reveals that for the same aspect ratio, the intercept is smaller for  Mn$_{12}$-ac-MeOH than it is in Mn$_{12}$-ac.

\begin{figure}[h]
\includegraphics[width=0.50\textwidth]{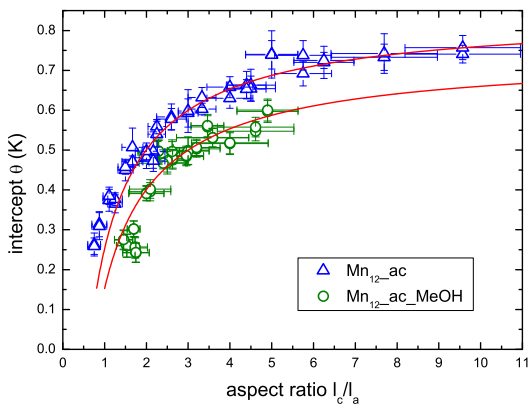}
\caption{\label{comparison}(Color online) Temperature intercept $\theta$ as a function
of aspect ratio for Mn$_{12}$-ac (triangles) and Mn$_{12}$-ac-MeOH (circles).  The lines denote fits obtained as described in the text.}
\end{figure}

The intercept $\theta$ is shown in Fig. \ref{comparison} as a function of aspect ratio for Mn$_{12}$-ac and Mn$_{12}$-ac-MeOH.  For each material, $\theta$ increases with increasing aspect ratio, asymptotically approaching a limiting value as the sample becomes longer and/or thinner.  While the behavior as a function of aspect ratio is qualitatively similar for the two materials, it is clear that $\theta$ is smaller in Mn$_{12}$-ac-MeOH than in Mn$_{12}$-ac for every aspect ratio over the entire range of our measurements.

\section{Theoretical Analysis}

The study of demagnetization factors of homogeneous bodies has been
a classical topic in magnetism \cite{Cullity,Chen1991}.
Demagnetization factors have been calculated for many different
shapes, including ellipsoids and spheres, rods and disks, and
rectangular prisms \cite{Chen1991,Chen2006,Chen2005,Chen2004}.  In
the analysis described below, we use published tables of the
demagnetization factors for bars with square cross-section
\cite{Chen2004,Chen2005} to fit the data of Fig. \ref{comparison} and determine the limiting value of $\theta_{CW} = T_c$ for very large aspect ratio.

The magnetic susceptibility measured in our experiment, $\chi=M/H_{ext}$, is deduced from the slope of the straight lines of $M$ versus the externally applied magnetic field $H_{ext}$ shown in Fig. \ref{hys}.  To obtain the true susceptibility, $\chi= M/H_{tot}$, one needs to use the total magnetic field, $H_{tot}=H_{ext}+H_d$, where $H_d$ is the demagnetizing field.  The demagnetizing field is opposite to and proportional to the magnetization of the sample, $H_d = -N_dM$, where $N_d$ depends primarily on the geometry of the sample and, to a lesser degree, the susceptibility of the material (see Chen {\it et al.} \cite{Chen1991} and references therein).  Except for ellipsoidal specimens, the demagnetization factor varies from point to point, and one needs to apply an averaged demagnetization factor that depends on the type of measurement - {\it e. g.} a measurement taken by a coil wound around the middle or a measurement of the entire sample.  For our SQUID-based measurements of small samples, the appropriate factor is the ratio of the average demagnetizing field to the average magnetization of the entire sample, the so-called magnetometric demagnetization factor $N_m$.

\begin{figure}[h]
\includegraphics[width=0.50\textwidth]{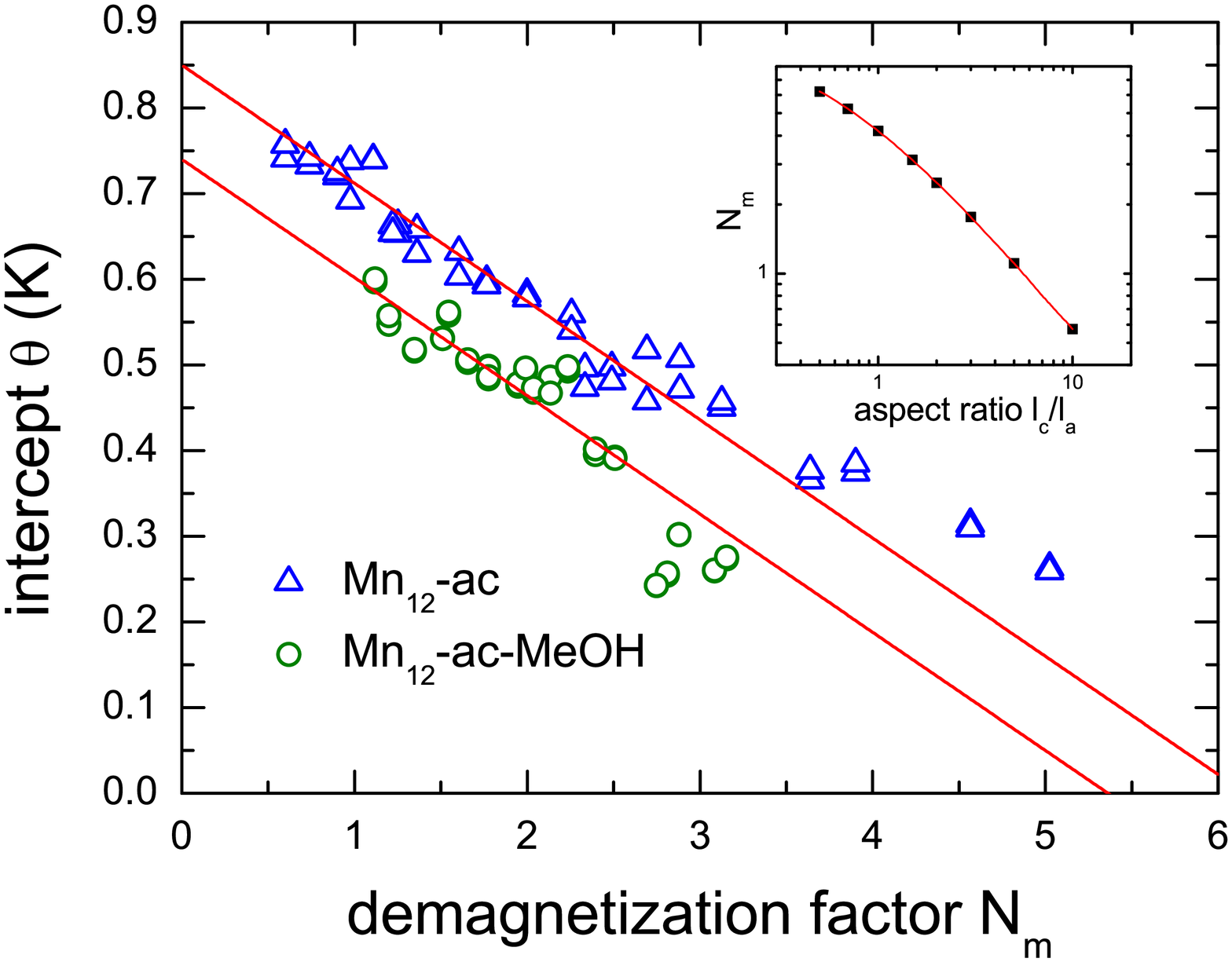}
\caption{\label{analysisnew} (Color online) Temperature intercept $\theta$ as a
function of demagnetization factor $N_m$ (in cgs units) for Mn$_{12}$-ac
(open circles) and Mn$_{12}$-ac-MeOH (closed dots).  Approximate fits are denoted by the solid lines with (negative) slopes constrained to be equal to the Curie constant $C=0.138$ (see Eq. \ref{final}).
Inset: Magnetometric demagnetization factor $N_m$ as a function of
aspect ratio; values of $N_m$ are obtained by interpolation from the
tables published by Chen, Pardo  and Sanchez
\cite{Chen2004,Chen2005}.}
\end{figure}

The magnetometric demagnetization factor, $N_m$, was obtained for
the aspect ratios of our crystals by interpolation using the
published tables for bars of square crossection \cite{Chen2004,Chen2005}.  We select
the values listed for $\chi=0$ since the small susceptibility of our
samples produces demagnetizing fields that are small compared to the
applied magnetic field.  The resulting curve for $N_m$ versus aspect
ratio is shown in the inset to Fig. \ref{analysisnew}.  Combining this with the information in Fig. \ref{comparison}, one obtains $\theta$ versus $N_m$ shown in the main part of the figure for Mn$_{12}$ and Mn$_{12}$-ac-MeOH.

The simplest mean field derivation of the Curie Weiss law incorporates the effect of interactions by postulating a ``molecular field'', $H_m$.  The demagnetizing field $H_d$ can be introduced in a similar way by writing $H_{tot}=H_{ext}+H_m+H_d$, from which one obtains:

\begin{equation}\label{final}
\theta=\theta_{CW}(\frac{c}{a})-C N_m (\frac{l_c}{l_a}),
\end{equation}
where $C$ is the Curie constant.  The first term in this expression depends only on lattice properties such as $c/a$ and the local chemistry (the molecule solvent and ligand structure)  and is independent of the aspect ratio, while the second term depends only on the shape of the crystal and vanishes in the
limit of infinite aspect ratio $l_c/l_a$.  For a particular material, say, Mn$_{12}$-ac, the lattice properties such as $c/a$ and local chemistry are the same for all samples with different aspect ratios, and the value of $\theta_{CW}$ in Eq.(\ref{final}) can be interpreted empirically as the intrinsic Curie-Weiss temperature $T_c$ obtained in the limit of infinite aspect ratio \cite{assumption}.

As shown in Fig. \ref{analysisnew}, $\theta$ depends linearly on $N_m$, as expected from Eq. \ref{final}, for both Mn$_{12}$-ac and Mn$_{12}$-ac-MeOH.  Guided by Eq. \ref{final}, the slopes of the solid lines drawn in the figure were constrained to the value $C=0.138$ obtained from the data of Fig. \ref{slope}, yielding $\theta_{CW} = T_c \approx 0.85 $ K for Mn$_{12}$-ac and $T_c\approx 0.74$ K for Mn$_{12}$-ac-MeOH.

Theoretical calculations \cite{Chudnovsky,Garanin,Millis} have been carried out for Mn$_{12}$-ac based on models that consider dipolar interactions only, on the assumption that other terms (for example, direct exchange from overlap of wavefunctions) can be neglected.  Chudnovsky and Garanin \cite{Chudnovsky} predicted ferromagnetic ordering of elongated crystals of Mn$_{12}$-ac below  $0.8$ K; Garanin's \cite{Garanin} recent investigation of elongated box-shape crystals yielded an ordering temperature  $\sim 0.71$ K.   Values of $J(c/a)=\theta_{CW}$ can also be obtained from the work of Millis {\it et al.} \cite{Millis}, who write the susceptibility as:
\begin{equation}\label{chi2}
\frac{1}{\chi}=\frac{T-\theta_{CW}}{C} = \frac{T-2E_{dip} J}{C}; J = J_{SR} + 4\pi/3
\end{equation}
Here $E_{dip}$ is the dipolar interaction and the short-range contribution $J_{SR}$, depends on the details of the crystal structure.  For Mn$_{12}$-ac, with lattice constants $a = b = 17.1668(3)$ \AA, $c = 12.2545(3)$ \AA, one obtains $J \approx 5.287$ while for Mn$_{12}$-ac-MeOH with $a = b = 17.3500(18)$ \AA, $c = 11.9971(17)$, $J \approx 5.514$.
The strength of the dipolar interaction, $E_{dip} \approx 0.078$ K, is essentially the same for the two materials as their unit cells have the same volume within $0.01$ \%.  This yields $\theta_{CW}  \approx 0.82$ K for Mn$_{12}$-ac and $\approx 0.86$ K for Mn$_{12}$-ac-MeOH.

The crystal structures of Mn$_{12}$-ac-MeOH and Mn$_{12}$-ac are quite similar: the unit cell parameters and unit volumes cell are nearly identical, and the strength of the dipolar interactions are expected to be essentially the same.  These similarities are reflected by the nearly identical Curie constants obtained experimentally for the two systems.  By contrast, however, the values of the Curie-Weiss $\theta_{CW}$'s are clearly and consistently smaller for Mn$_{12}$-ac-MeOH than they are in Mn$_{12}$-ac, as shown in Figs. \ref{comparison} and \ref{analysisnew}, implying that the magnetic interactions are weaker in Mn$_{12}$-ac-MeOH.  We note that although the unit cell parameters are nearly identical, the two systems have different ligands bridging the Mn$_{12}$ molecules.  A possible explanation for the different interaction strengths in the two materials may be that, in addition to the dipolar interactions, quantum mechanical exchange deriving from wave function overlap plays a significant role.  In particular, our results suggest that there is an extra direct exchange contribution of antiferromagnetic sign in Mn$_{12}$-ac-MeOH.

\vspace{0.06in}

\section{Summary}
The susceptibility of Mn$_{12}$-ac and Mn$_{12}$-ac-MeOH has been measured for a series of samples in the shape of rectangular prisms
of length $l_c$ and square cross-section of side $l_a$.  Fits to a Curie-Weiss Law, $\chi=C/(T-\theta)$, yield values for $\theta$ that vary systematically with the aspect ratio, $l_c/l_a$.   Using published values of the demagnetization factor \cite{Chen2004,Chen2005} we have deduced values of $\theta_{CW}=T_c$ that are surprisingly different for Mn$_{12}$-ac and Mn$_{12}$-ac-MeOH, two materials that have nearly identical crystal structures but different ligands bridging the Mn$_{12}$ molecules in the crystal.  This suggests that, in addition to dipolar interactions, there is a non-dipolar (exchange) contribution to the Weiss temperature that is different for the different ligand molecules in the crystal.

\section{Acknowledgments}

We thank Dimitar Dimitrov for valuable technical help during the initial phases of the experiment. We acknowledge illuminating discussions with D. M. Garanin and E. M. Chudnovsky.  Support for GC was provided by NSF under grant
CHE-0910472; ADK acknowledges support by NSF-DMR-0506946 and ARO
W911NF-08-1-0364; AJM acknowledges support of NSF-DMR-0705847; MPS
acknowledges support from NSF-DMR-0451605; YY acknowledges support of the
Deutsche Forschungsgemeinschaft through a Deutsch-Israelische
Projektkooperation (DIP).


\begin{thebibliography}{99}
\bibitem{comment} Samples of widely different size with the same aspect ratio yielded closely similar results, confirming the validity of our analysis.
\bibitem{assumption} We assume the relative difference in $c/a$ does not change with temperature.
\bibitem{Garanin} D. A. Garanin, Phys. Rev. B {\bf 81}, 220408 (2010); note that Garanin claims in this paper that finding the ordering temperature by linearly extrapolating the inverse susceptibility curve above the transition temperature does not provide a correct $T_c$.
\end{thebibliography}
\end{document}